\documentclass{TTP_DSL2006}
 
\usepackage[dvipsone]{graphicx}
\usepackage[intlimits]{amsmath}
\usepackage{amssymb}
\usepackage{exscale}
\usepackage[colorlinks,urlcolor=blue]{hyperref}

\usepackage{sidecap}

\begin{document}

\title{Charge carrier localisation in disordered graphene nanoribbons}

\author{ Gerald Schubert\inst{1,2}
%$^{,\rm{a}}$, 
%Oleg P. Shuskov\inst{3}, 
and Holger Fehske\inst{1,3}}
%$^{,\rm{b}}$ }

\institute{
Institut f{\"u}r Physik, Ernst-Moritz-Arndt-Universit{\"a}t Greifswald, 17487 Greifswald, Germany
\and
Regionales Rechenzentrum Erlangen, Universit{\"a}t Erlangen-N{\"u}rnberg, 91058 Erlangen, Germany
\and
School of Physics, University of New South Wales, Kensington 2052, Sydney NSW, Australia
}
\maketitle
  
%\vspace{-3mm}
%\sffamily
%\begin{center}
%$^{a}$gerald.schubert@rrze.uni-erlangen.de, $^{ b}$holger.fehske@physik.uni-greifswald.de 
%\end{center}

\vspace{2mm} \hspace{-7.7mm} \normalsize \textbf{Keywords:} graphene, local density of states, disorder effects, metal-insulator transition\\

\vspace{-2mm} \hspace{-7.7mm}
\rmfamily
\noindent \textbf{Abstract.} 
We study the electronic properties of actual-size graphene nanoribbons
subjected to substitutional disorder particularly with regard to the
experimentally observed metal-insulator transition. Calculating the
local, mean and typical density of states, as well as the
time-evolution of the particle density we comment on a possible
disorder-induced localisation of charge carriers at and close to the
Dirac point within a percolation transition scenario.

\section{Introduction}
Transport in graphene nanoribbons (GNRs) is strongly affected by
disorder effects which can be traced back, for example, to
dislocations or charged impurities in the substrate, to adatoms
adsorbed at the graphene surface, to edge defects, or to ripples
associated with the soft structure of graphene.
When modelling disorder, many theoretical studies resort to the
generic Anderson model~\cite{An58}, which exhibits a disorder-induced
localisation transition in three dimensions (3D) that is absent in
lower dimensions, however.  One-parameter scaling theory predicts that
all states are localised for the infinite Anderson-disordered 2D
system~\cite{AALR79}.
The recently observed metal-insulator transition in hydrogenated
graphene~\cite{BMESHKR09}, disordered GNRs~\cite{ACFD08}, and
Si-MOSFET inversion layers~\cite{KKFPD94c} is beyond reach of the
Anderson model, but might be explainable by percolation-based
approaches~\cite{DLHPWR05,ACFD08}.

The classical (geometric) problem of percolation consists in finding a
connected path of accessible sites that spans the whole lattice.
On the honeycomb lattice, for site-percolation, this will be the case
for a concentration of accessible sites
$p>p_c\approx0.697$~\cite{SZ99}.
For real materials the strict distinction between accessible and
blocked sites seems to be too simplistic.
For example, upon hydrogenation, the $\pi$-bonds of some carbon atoms
within a graphene sheet will be blocked just partly.  Also the
electron-hole puddles resulting from charged impurities in the
substrate lead to a finite difference between the on-site potentials
only.
Then, tunneling effects between the puddles may become possible,
allowing for transport despite the absence of a percolating cluster.
In the quantum case, a spanning cluster does not guarantee transport
since scattering at its irregular boundaries causes interference
effects which may lead to localisation of the charge carriers.

To analyse the localisation properties of low-dimensional systems
within numerical approaches, both sophisticated algorithms and highly
efficient implementations are mandatory since the relevant length
scales are exceptionally large.
In this regard the finite extension of mesoscopic graphene flakes or
GNRs deserves further attention since it may mask localisation effects
if the localisation length exceeds the system size.%

In this work we investigate the electronic properties of GNR
eigenstates in the vicinity of the Dirac point using a local
distribution approach based on exact diagonalisation (ED)
techniques~\cite{AF08}.  This method has proven its reliability for
the study of Anderson localisation and quantum percolation on various
lattices in different dimensions.~\cite{SF08,SSBFV10}
 
\section{Model and method}
To model disordered GNRs we consider the tight-binding Hamiltonian
\begin{equation}
  \label{model}
  {H} =  - \bar{t} \sum\limits_{\langle ij \rangle} 
  \bigl({c}_i^{\dag} {c}_j^{} + \text{H.c.}\bigr)
  + \sum\limits_{j=1}^{N} \epsilon_j {c}_j^{\dag} {c}_j^{} 
\end{equation}
on a honeycomb lattice with $N$ sites, including electron transfer
$\bar{t}$ between nearest neighbours $\langle ij\rangle$ only.  In
Eq. (\ref{model}), $c_i^{\dag}$ $(c_i)$ creates (annihilates) an
electron at lattice site $i$.  Drawing the on-site potentials
$\epsilon_i$ from the bimodal distribution
\begin{equation}
  p[\epsilon_j] = p\,\delta(\epsilon_j-\epsilon_A) +
  (1-p)\, \delta(\epsilon_j-\epsilon_B)\;,
  \label{binalloy}
\end{equation}
sites are occupied by an atom of type $A$ [$B$] with probability $p$
[$(1-p)$] (binary alloy analogy).  Without loss of generality we
choose the on-site energy of the majority sub-band $\epsilon_A=0$, use
$\Delta=\epsilon_B-\epsilon_A$ hereafter, and let $\bar{t}$ fix the
unit of energy.

Localisation properties of disordered systems can be discussed in
terms of the local density of states (LDOS),
\begin{equation} \label{LDOS}
  \rho_i(E) = \sum\limits_{n=1}^{N}
  | \langle i | n \rangle |^2\, \delta(E-E_n)\;,
\end{equation}
where $|i\rangle = c^{\dag}_i |0\rangle$, and $|n\rangle$
is a single-electron eigenstate of $H$ with energy $E_n$.
The LDOS can be determined very efficiently by the Kernel Polynomial
Method which is an expansion of the rescaled Hamiltonian into a finite
series of Chebyshev polynomials~\cite{WWAF06}.  Thereby an energy
level broadening appears that can be controlled by the expansion
order.

Within the local distribution approach one has to analyse the
behaviour of the normalised LDOS distribution
$f[\rho_i/\rho_{\text{me}}]$ upon finite-size scaling for many
realisations of disorder (here $\rho_{\text{me}}=\langle
\rho_i\rangle$ is the mean DOS). While extended states are
characterised by a system-size independent
$f[\rho_i/\rho_{\text{me}}]$, the distribution for localised states
strongly depends on $N$; its maximum shifts towards zero and finally
$f[\rho_i/\rho_{\text{me}}]$ becomes singular as $N\to\infty$.
Since in many cases $f[\rho_i/\rho_{\text{me}}]$ closely resembles a
log-normal distribution~\cite{SSBFV10}, a simplified discussion may
use the typical DOS, $\rho_{\text{ty}}=e^{\langle\ln(\rho_i)\rangle}$,
which is directly related to the maximum of the log-normal
distribution.

Alternatively we may access the localisation properties of a system by
determining the recurrence probability $P_R(t\to\infty)$, which in the
thermodynamic limit is finite for localised states and scales to zero
as $1/N$ for extended states.  Again a finite series expansion into
Chebyshev polynomials is promising, in this case applied to the time
evolution operator~\cite{SF08}. %\cite{TK84,SF08,FSSWFB09}.
Then, for example, we may track how an initially localised wave packet
evolves in time by calculating the time dependent local particle
density,
\begin{equation}
  |\psi({\mathbf r}_i, t)|^2 = \Big| \sum\limits_{m=1}^{N}  
  e^{-{\mathrm i} E_m t} \langle m| \psi(0)\rangle \langle i| m \rangle \Big|^2\;.
\end{equation}

\section{Numerical results}

%
%%%%%%%%%%%%%%%%%%%%%%%%%%%%%%%%%%%%%%%%%%%%%%%%%%%%%%%%%%%%%%%%%%%%%%%%%
%
\begin{SCfigure}\centering
 \includegraphics[width=0.65\linewidth,clip]{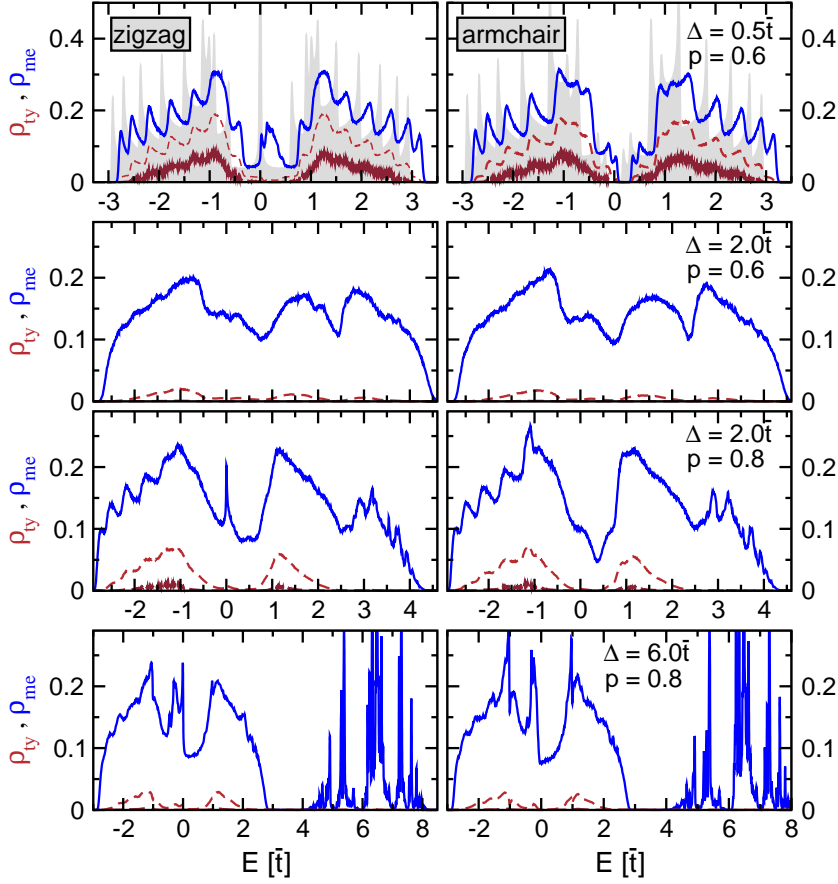}
 \caption{\label{fig:metyDOS}Mean (solid blue) and typical (dashed
   red) DOS for zigzag (left column, $N_z=6$) and armchair (right
   column, $N_a=10$) GNRs of width $W=1.1\,\mathrm{nm}$ and periodic
   boundary conditions in longitudinal direction. In each panel
   $\rho_{\text{ty}}$ is given for $L = 213\,(1064)\,\mathrm{nm}$ by
   red dashed (dark-red long-dashed) lines for $6\times10^4$
   realisations of disorder.  These system sizes correspond to $10000$
   $(50000)$ lattice sites for the armchair and $10392$ $(51960)$ for
   the zigzag case. For comparison the mean DOS for pristine GNRs is
   given in grey in the top panels.
   %\textcolor{red}{KPM resolution: $N\sigma=2.5$}
 }
\end{SCfigure}
%
%%%%%%%%%%%%%%%%%%%%%%%%%%
%
\begin{SCfigure}\centering
 \includegraphics[width=0.65\linewidth,clip]{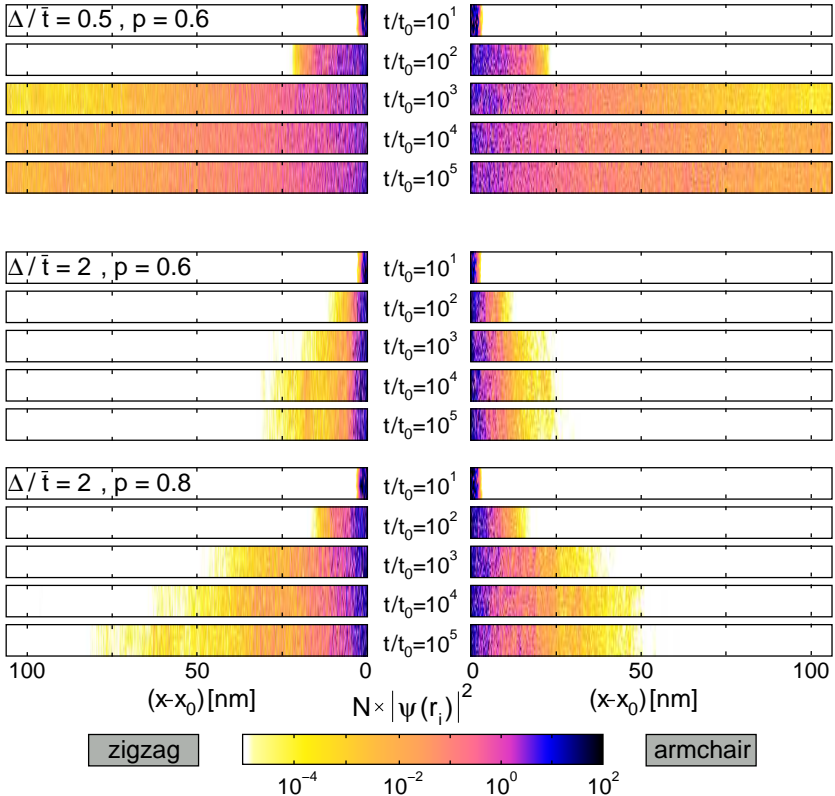}
 \caption{\label{fig:TimeEv}Time evolution of the normalised particle
   density $N |\psi({\mathbf r}_i)|^2$ for disordered GNRs. Device
   dimensions (only one half is shown): $(1.1\times213)\,\text{nm}^2$
   corresponding to $6\times 1732$ atoms (zigzag) and $10\times 1000$
   atoms (armchair). Times are measured in units of the inverse
   hopping element $t_0=1/\bar{t}$.}
\end{SCfigure}
%
%%%%%%%%%%%%%%%%%%%%%%%%%%
%
\begin{SCfigure}\centering
 \includegraphics[width=0.6\linewidth,clip]{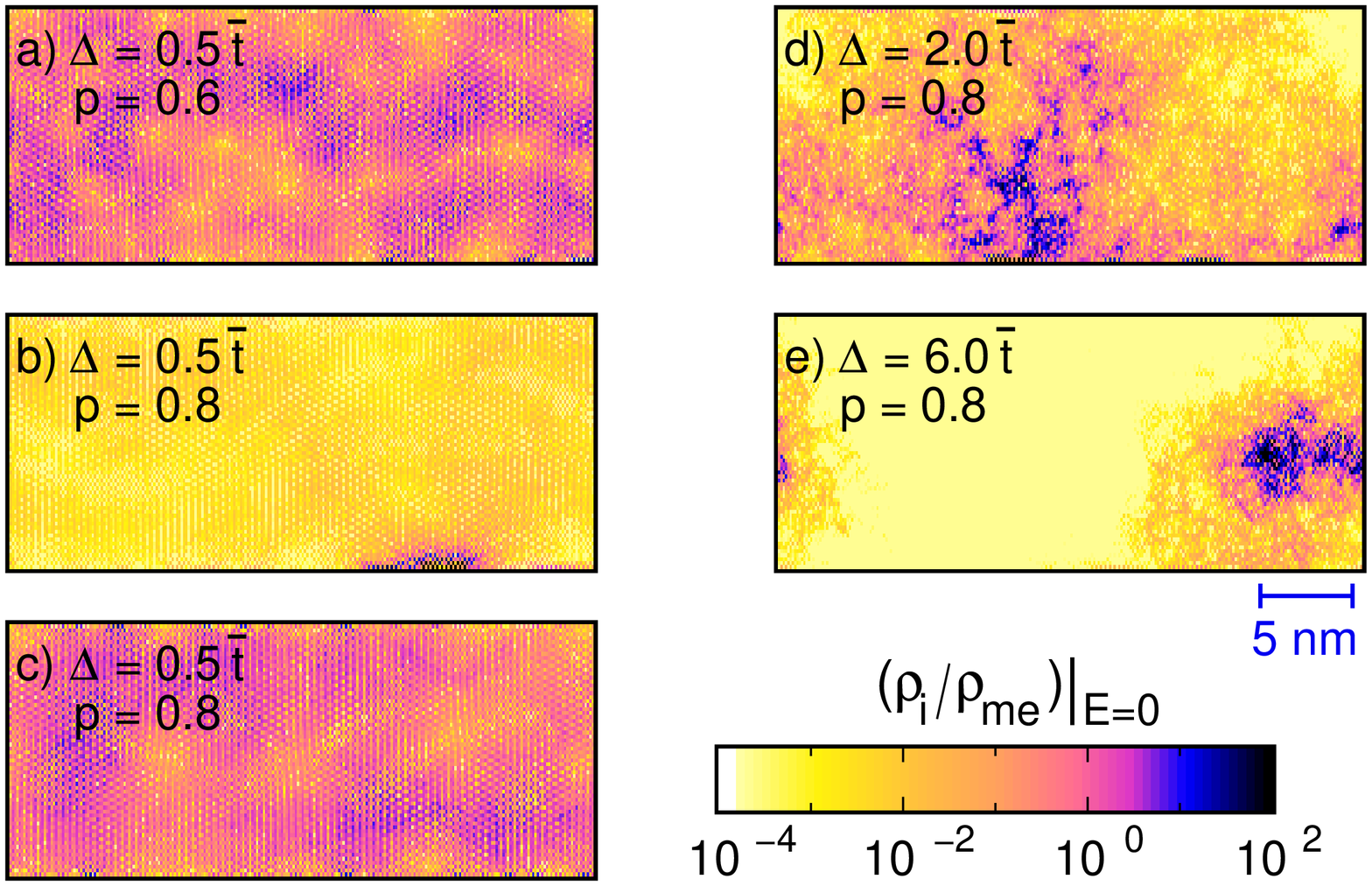}
 \caption{\label{fig:CharState} Normalised LDOS
   $(\rho_i/\rho_{\text{me}})|_{E=0}$ for particular zigzag GNRs with
   $256\times64$ sites as obtained by exact diagonalisation. In
   contrast to the other panels, where the LDOS is calculated at the
   unperturbed Dirac point, $E=0$, panel (c) refers to
   $E^{-}=-0.007\bar{t}$.}
\end{SCfigure}

As compared to mesoscopic graphene flakes, the DOS of ordered GNRs
exhibits a series of spikes (van Hove singularities) indicating
quasi-1D behaviour. The number and position of these spikes depend on
the ribbon geometry and on the (finite) number of unit cells in the
transverse cross section.  Furthermore, GNRs form very interesting
edge states. While for armchair edges there is a gap at the Dirac
point, $E=0$, a variety of degenerate edge states exist for zigzag
edges (see top panel of Fig.~\ref{fig:metyDOS}).

Accounting for adatoms that lead to random on-site potentials
$\Delta$, a ``copy'' of the DOS centred around $\Delta$ will appear.
The random superposition of the local A- and B-atom DOS results in the
total (mean) DOS presented in Fig.~\ref{fig:metyDOS}.
Increasing the value of $\Delta$ enhances the random on-site
fluctuations, and at $\Delta=6\bar{t}$ (bottom panel of
Fig.~\ref{fig:metyDOS}) we enter a split-band regime, where two
distinct sub-bands emerge.

The decay of $\rho_{\text{ty}}$ with increasing system size indicates
that the states are localised in principle for any of the shown
parameters; actually $\rho_{\text{ty}}$ vanishes for
$N\sim5\times10^4$ already in all panels except for those in the top
most row. There, however, the finite value of $\rho_{\text{ty}}$
solely indicates that the localisation length is larger than the
system size or comparable to it.

%
%%%%%%%%%%%%%%%%%%%%%%%%%%%%%%%%%%%%%%%%%%%%%%%%%%%%%%%%%%%%%%%%%%%%%%%%%
%

Figure~\ref{fig:TimeEv} shows the time evolution of an initially
localised state, as calculated by the Chebyshev method.
After an initial, fast spreading process ($t\lesssim10^4t_0$) the wave
function becomes quasi-stationary, i.e., there are temporal amplitude
fluctuations on individual sites but the overall region of sites
having finite amplitudes remains constant.
Moderate ribbon lengths and small values of $\Delta$ result in a
localisation length larger than the system size (top panel of
Fig.~\ref{fig:TimeEv}) and therefore cause a ``metallic'' behaviour of
the GNR.
In contrast, for larger $\Delta$ the wave packet remains localised
even for rather small systems.
Clearly the wave function stays more localised for $p=0.6<p_c$ than
for $p=0.8>p_c$, since due to the lack of a spanning cluster in the
first case, the spreading of the wave function depends heavily on
quantum tunneling, whose efficiency decreases as $\Delta$ increases.
%
%%%%%%%%%%%%%%%%%%%%%%%%%%%%%%%%%%%%%%%%%%%%%%%%%%%%%%%%%%%%%%%%%%%%%%%%%
%

To illustrate the localisation properties of our binary-alloy GNR
model in more detail, we present ED results for the normalised LDOS in
Fig.~\ref{fig:CharState}, focusing thereby on energies close to the
(unperturbed) Dirac point, $E=0$.
Although for a small potential difference $\Delta$ and a nearly equal
concentration of A and B atoms the wave function spans the entire
(finite) lattice, the bimodal distribution still shows up by two
interpenetrating regions with large and small wave-function amplitudes
[see Fig.~\ref{fig:CharState}\,(a)]. A stronger asymmetry in the
concentration of type A and B atoms then leads to a localisation of
the wave function in small regions near the GNR's edge
[cf. Fig.~\ref{fig:CharState}\,(b)].
As may be expected the wave-function localisation effect becomes more
pronounced at larger $\Delta$, i.e. the localisation length clearly
decreases upon increasing $\Delta$ [see Fig.~\ref{fig:CharState}\,(d)
and (e)].  Note, however, that for $p=0.8$ a spanning cluster exists,
that is we observe a true quantum percolation effect.

A particularly interesting feature is the complete change in character
of the eigenstates when crossing the energy of the unperturbed Dirac
point for slightly differing on-site potentials and asymmetric
distribution of the atom variants.
While for $E=0^{+}$ [Fig.~\ref{fig:CharState}\,(b)], the state is
clearly localised, states on the opposite side of the impurity
sub-band, $E=0^{-}$, are extended [Fig.~\ref{fig:CharState}\,(c)].
This transition is absent for a less pronounced asymmetry between the
two atom types, e.g. for $p=0.6$ in [Fig.~\ref{fig:CharState}\,(a)].
%
%%%%%%%%%%%%%%%%%%%%%%%%%%%%%%%%%%%%%%%%%%%%%%%%%%%%%%%%%%%%%%%%%%%%%%%%%
%
%\section*{Conclusions}

In conclusion, we have argued theoretically and demonstrated
numerically by zero-tempera-ture exact diagonalisation calculations
that disorder of binary-alloy type in combination with quantum
percolation effects will strongly affect electron transport in
graphene nanoribbons, up to the point of a disorder-induced
localisation of the charge carriers.  We argue that in order to
corroborate such kind of quantum localisation, transport measurements
should be performed at much lower temperatures than used so far, but
even so the effect might be covered by the large localisation lengths
compared to the spatial dimensions of actual GNRs.

%
%%%%%%%%%%%%%%%%%%%%%%%%%%%%%%%%%%%%%%%%%%%%%%%%%%%%%%%%%%%%%%%%%%%%%%%%%
%
\section*{Acknowledgements}
This work was funded by the Competence Network for
Technical/Scientific High-Performance Computing in Bavaria (KONWIHR)
and the Deutsche Forschungsgemeinschaft through Priority Program SPP
1459. HF acknowledges the hospitality at the Massey University
Palmerston North (New Zealand) and the Gordon Godfrey visiting
fellowship by the UNSW (Australia).
%
%%%%%%%%%%%%%%%%%%%%%%%%%%%%%%%%%%%%%%%%%%%%%%%%%%%%%%%%%%%%%%%%%%%%%%%%%
%

%\bibliography{./ref} 

%\bibliography{./sf11} 

\begin{thebibliography}{10}

\bibitem{An58} P.~W. Anderson.
  \newblock {\em Phys. Rev.}, 109:1492, 1958.

\bibitem{AALR79} E. Abrahams, P.~W. Anderson, D. C. Licciardello, and T. V. Ramakrishnan.
  \newblock {\em Phys. Rev. Lett.}, 42:673, 1979.

\bibitem{BMESHKR09} A. Bostwick, J.~L. McChesney, K.~V. Emtsev, T. Seyller, K. Horn, S.~D. Kevan, and E. Rothenberg.
  \newblock {\em Phys. Rev. Lett.}, 103:056404, 2009.

\bibitem{ACFD08} S.~Adam, S.~Cho, M.~S. Fuhrer, and S.~Das~Sarma.
  \newblock {\em Phys. Rev. Lett.}, 101:046404, 2008.

\bibitem{KKFPD94c} S.~V. Kravchenko, G.~V. Kravchenko, J.~E. Furneaux, V.~M. Pudalov, and M.~D'Iorio.
  \newblock {\em Phys. Rev. B}, 50:8039, 1994;
  \newblock S.~V. Kravchenko, W.~E. Mason, G.~E. Bowker, J.~E. Furneaux, V.~M. Pudalov, and M.~D'Iorio.
  \newblock {\em Phys. Rev. B}, 51:7038, 1995.

\bibitem{DLHPWR05} S.~Das~Sarma, M.~P. Lilly, E.~H. Hwang, L.~N. Pfeiffer, K.~W. West, and J.~L. Reno.
  \newblock {\em Phys. Rev. Lett.}, 94:136401, 2005.
  
\bibitem{SZ99}P.~N.~Suding and R.~M.~Ziff.
  \newblock {\em Phys. Rev. E}, 60:275, 1999.

\bibitem{AF08} A. Alvermann and H. Fehske.
\newblock {\em Lect. Notes Phys.}, 739:505, 1999.

\bibitem{SF08} G. Schubert and H. Fehske.
  \newblock {\em Phys. Rev. B}, 77:245130, 2008;
  \newblock G. Schubert, J. Schleede and H. Fehske.
  \newblock {\em Phys. Rev. B}, 79:235116, 2009.

\bibitem{SSBFV10} G. Schubert, J. Schleede, K. Byczuk, H. Fehske, and D. Vollhardt.
  \newblock {\em Phys. Rev. B}, 81:155106, 2010.

\bibitem{WWAF06} A. Wei{\ss}e, G. Wellein, A. Alvermann, and H. Fehske.
  \newblock {\em Rev. Mod. Phys.}, 78:275--306, 2006.

\end{thebibliography}
%\bibliographystyle{plain}

\end{document}